\documentclass[conference]{IEEEtran}

\usepackage{amsmath}
\usepackage{amsthm}
\usepackage{amssymb}

\usepackage{cite}
\usepackage{graphicx}

\newtheorem{lemma}{Lemma}
\newtheorem{observation}{Observation}

\begin{document}

\newcommand{\mathacr}[1]{\mathsf{#1}}
\newcommand{\argmax}[1]{{\underset{{#1}}{\mathrm{arg\,max}}}}
\newcommand{\argmin}[1]{{\underset{{#1}}{\mathrm{arg\,min}}}}
\newcommand{\vect}[1]{\mathbf{#1}}
\newcommand{\maximize}[1]{{\underset{{#1}}{\mathrm{maximize}}}}
\newcommand{\minimize}[1]{{\underset{{#1}}{\mathrm{minimize}}}}
\newcommand{\condSum}[3]{\overset{#3}{\underset{\underset{#2}{#1}}{\sum}}}
\newcommand{\condProd}[3]{\overset{#3}{\underset{\underset{#2}{#1}}{\prod}}}

\def\Psiv{\vect{\Psi}}
\def\diag{\mathrm{diag}}
\def\eig{\mathrm{eig}}
\def\kron{\otimes}
\def\tr{\mathrm{tr}}
\def\rank{\mathrm{rank}}
\def\Htran{\mbox{\tiny $\mathrm{H}$}}
\def\Ttran{\mbox{\tiny $\mathrm{T}$}}
\def\CN{\mathcal{N}_{\mathbb{C}}} 
\def\taupu{\tau_{p}} 
\def\taupd{\tau_{p}^{d}} 
\def\bphiu{\boldsymbol{\phi}} 
\def\bphid{\boldsymbol{\varphi}} 
\def\Ktot{K_{\mathrm{tot}}}
\def\Pu{\mathcal{P}} 
\def\sigmaDL{\sigma^2_{\mathrm{DL}}} 
\def\sigmaUL{\sigma^2_{\mathrm{UL}}} 
\def\kappatUE{\kappa_t^{\mathrm{UE}}} 
\def\kapparUE{\kappa_r^{\mathrm{UE}}} 
\def\kappatBS{\kappa_t^{\mathrm{BS}}} 
\def\kapparBS{\kappa_r^{\mathrm{BS}}} 
\def\ple{\alpha} 
\def\imagunit{\mathsf{j}} 
\def\Punder{\underline{P}}
\def\Pover{\overline{P}}
\def\EcondH{\mathbb{E}_{| \vect{H}}}
\def\boff{b_{\mathrm{off}}}

\title{Can Hardware Distortion Correlation be Neglected When Analyzing Uplink SE in Massive MIMO?\vspace{-.3cm}}

\author{
\IEEEauthorblockN{Emil Bj{\"o}rnson\IEEEauthorrefmark{1},
Luca Sanguinetti\IEEEauthorrefmark{2}, Jakob Hoydis\IEEEauthorrefmark{3} \bigskip
\thanks{This research has been supported by ELLIIT, the EU FP7 under ICT-619086 (MAMMOET) and the ERC Starting Grant 305123 MORE.}
\vspace{-.3cm}}
\IEEEauthorblockA{\IEEEauthorrefmark{1}\small{Department of Electrical Engineering (ISY), Link\"{o}ping University, Link\"{o}ping, Sweden}}
\IEEEauthorblockA{\IEEEauthorrefmark{2}\small{Dipartimento di Ingegneria dell'Informazione, University of Pisa, Pisa, Italy}}
\IEEEauthorblockA{\IEEEauthorrefmark{3}\small{Nokia Bell Labs, Nozay, France.}}
\vspace{-.8cm}}

\maketitle

\begin{abstract}
This paper analyzes how the distortion created by hardware impairments in a multiple-antenna base station affects the uplink spectral efficiency (SE), with focus on Massive MIMO. The distortion is correlated across the antennas, but has been often approximated as uncorrelated to facilitate (tractable) SE analysis. To determine when this approximation is accurate, basic properties of the distortion correlation are first uncovered. Then, we focus on third-order non-linearities and prove analytically and numerically that the correlation can be neglected in the SE analysis when there are many users. In i.i.d.~Rayleigh fading with equal signal-to-noise ratios, this occurs when having five users.
\end{abstract}


\IEEEpeerreviewmaketitle

\section{Introduction}

The receiver hardware in a wireless communication system is always non-ideal \cite{Schenk2008a}; for example, there are non-linear amplification in the low-noise amplifier (LNA) and quantization errors in the analog-to-digital converter (ADC). In single-antenna receivers, these effects can be equivalently represented by scaling the received signal and adding uncorrelated (but not independent) noise \cite{Bussgang1952a,Fletcher2007a}.
This is a convenient representation in information theory, since one can compute lower bounds on the capacity by utilizing that the worst-case uncorrelated additive noise is independent and Gaussian distributed~\cite{Hassibi2003a}.

The extension to multiple-antenna receivers is non-trivial, but important since Massive MIMO (multiple-input multiple-output) is the key technology to improve SE in future networks. In Massive MIMO, a base station (BS) with $M\geq 100$ antennas is typically used to serve $K\geq 10$ user equipments (UEs) by spatial multiplexing \cite{Marzetta2010a, BHS18A}. Each antenna is equipped with a separate transceiver chain (including LNA and ADC), so that one naturally assumes uncorrelated distortion between the antennas. This assumption was made in \cite{Bai2013a,Orhan2015a} (among others) without discussion, while the existence of correlation was mentioned in \cite{Bjornson2014a}, but claimed to be negligible in the SE analysis of Rayleigh fading Massive MIMO systems.

If two antennas equipped with identical hardware would receive identical signals, the distortions would be identical. Hence, it is important to characterize under which conditions, if any, the distortion can be reasonably modeled as uncorrelated across the receive antennas. 
In the related scenario of multi-antenna transmission, \cite{Moghadam2012a} conjectured that the radiated hardware distortion is ``\emph{practically uncorrelated}'' among antennas in MIMO systems transmitting multiple data streams. This conjecture holds when many data streams are transmitted with similar power \cite{Mollen2018b}.
Recently, \cite{Larsson2018a} claimed that the model with uncorrelated BS distortion is ``{\emph{physically inaccurate}}'' and \cite{Mollen2018a} stated that it ``{\emph{does not reveal the spatial characteristics of the distortion}}''. These papers primarily focus on out-of-band interference and assume ideal UE hardware.

The question is now if these statements are applicable or not when quantifying the SE in Rayleigh fading scenarios with both UE and BS impairments. This paper takes a close look at this issue, with the aim of providing a versatile view of when models with uncorrelated hardware distortion can be used for uplink  SE analysis with multiple-antenna BSs. To this end, we develop in Section~\ref{sec:system-model} a system model with a non-ideal multiple-antenna receiver and explain the characteristics of hardware distortion along the way. Achievable SE expressions are provided and maximized with respect to the receive combining. We study  amplitude-to-amplitude (AM-AM) non-linearities analytically in Section~\ref{sec:non-linearities}, with focus on maximum ratio (MR) combining. We demonstrate numerically when the distortion is well approximated as uncorrelated in SE analysis.

\vspace{-1mm}

\section{System Model and Spectral Efficiency}
\label{sec:system-model}

We consider the uplink of a single-cell system where $K$ single-antenna UEs communicate with a BS equipped with $M$ antennas. We consider a symbol-sampled complex baseband system model \cite{Mollen2018a}. The channel from UE $k$ is denoted by $\vect{h}_k =[h_{k1} \, \ldots \, h_{kM}]^{\Ttran} \in \mathbb{C}^{M}$. A block-fading model is considered where the channels are fixed within a time-frequency coherence block and take independent realizations in each block, according to an ergodic stationary  random process. In an arbitrary coherence block, the noise-free signal $\vect{u} = [u_1 \, \ldots \, u_M]^{\Ttran} \in \mathbb{C}^{M}$ received at the BS antennas is \vspace{-1mm}
\begin{equation}
\vect{u} = \sum_{k=1}^{K} \vect{h}_k s_k = \vect{H} \vect{s}
\end{equation} \vskip-2mm
\noindent where $\vect{H} = [\vect{h}_1 \, \ldots \, \vect{h}_{K}]\in \mathbb{C}^{M\times K}$ is the channel matrix, $s_k$ is the information signal transmitted by UE $k$, and $\vect{s}= [s_1 \, \ldots \, s_M]^{\Ttran}\sim \CN(\vect{0},p \vect{I}_K)$, where $\CN$ denotes the multi-variate circular symmetric complex Gaussian distribution.
We will analyze how $\vect{u}$ is affected by non-ideal receiver hardware and additive noise. To focus only on the distortion characteristics, $\vect{H}$ is assumed known.
\vspace{-2mm}
\subsection{Basic Modeling of BS Receiver Hardware Impairments}

We focus now on an arbitrary coherence block with the fixed channel realization $\vect{H}$ and use $\EcondH\{ \cdot \}$ to denote the conditional expectation given $\vect{H}$. Hence, the conditional distribution of $\vect{u}=[u_1 \, \ldots \, u_M]^{\Ttran}$ is  $\CN(\vect{0},\vect{C}_{uu})$ where $\vect{C}_{uu} = \EcondH\{\vect{u} \vect{u} ^{\Htran}\} = p \vect{H} \vect{H}^{\Htran} \in \mathbb{C}^{M \times M}$ describes the correlation between signals received at different antennas. It is only when $\vect{C}_{uu}$ is a scaled identity matrix that $\vect{u}$ has uncorrelated elements. This can only happen when $K \geq M$. However, we stress that $K < M$ is of main interest in Massive MIMO (or more generally in any multi-user MIMO system).

The BS hardware is assumed to be non-ideal but quasi-memoryless. The impairments at antenna $m$ are modeled by an arbitrary deterministic function $g_m(\cdot): \mathbb{C} \to \mathbb{C}$, for $m=1,\ldots,M$, which can describe both continuous non-linearities and discontinuous quantization errors.
These functions distort each of the components in $\vect{u}$, such that the resulting signal is
\begin{equation} \label{eq:z-distortion}
\vect{z} = \begin{bmatrix}
g_1(u_1) \, \ldots \, g_M(u_M)
\end{bmatrix}^{\Ttran} \triangleq \boldsymbol{g}(\vect{u}).
\end{equation}
By defining $\vect{C}_{zu} = \EcondH\{ \vect{z} \vect{u}^{\Htran} \}$, we can rewrite \eqref{eq:z-distortion} as
\begin{equation} \label{eq:z-distortion2}
\vect{z} = \vect{C}_{zu} \vect{C}_{uu}^{\dagger} \vect{u} + \boldsymbol{\eta}
\end{equation}
where we defined the additive distortion $\boldsymbol{\eta} \triangleq \boldsymbol{g}(\vect{u}) - \vect{C}_{zu} \vect{C}_{uu}^{\dagger} \vect{u}$ and $^\dagger$ denotes the pseudoinverse. 
By construction, $\vect{u} $ is uncorrelated with $\boldsymbol{\eta}$; that is, $\EcondH \{ \boldsymbol{\eta} \vect{u}^{\Htran} \} = \vect{C}_{zu} -  \vect{C}_{zu} \vect{C}_{uu}^{\dagger} \vect{C}_{uu} = \vect{0}$.
However, $\vect{u} $ and $\boldsymbol{\eta}$ are clearly not independent.

This derivation has not utilized the fact that $\vect{u}$ is Gaussian distributed (for a given $\vect{H}$), but only its first and second order moments. Hence, the model in \eqref{eq:z-distortion2} holds also for finite-sized constellations. Since the SE will be our metric, we utilize the full distribution to simplify $\vect{C}_{zu} \vect{C}_{uu}^{\dagger}$ in \eqref{eq:z-distortion2} using a discrete complex-valued counterpart to Bussgang's theorem \cite{Bussgang1952a}.

\begin{lemma} \label{lemma:bussgang}
Consider the jointly circular-symmetric complex Gaussian variables $x$ and $y$. For any deterministic function $f(\cdot): \mathbb{C}\to \mathbb{C}$, it holds that
\begin{equation} \label{eq:Bussgang-expression}
\mathbb{E}\{ f(x) y^*\} = \mathbb{E}\{ f(x) x^*\} \frac{\mathbb{E}\{ x y^*\}}{\mathbb{E}\{ |x|^2 \}}.
\end{equation}
\end{lemma}
\begin{IEEEproof}
Note that $y = \frac{\mathbb{E}\{ y x^*\}}{\mathbb{E}\{ |x|^2 \}} x + \epsilon$, where $\epsilon = y- \frac{\mathbb{E}\{ y x^*\}}{\mathbb{E}\{ |x|^2 \}} x$ has zero mean and is uncorrelated with $x$. The Gaussian distribution implies that $\epsilon$ and $x$ are independent. Replacing $y$ with this expression in the left-hand side of \eqref{eq:Bussgang-expression} and noting that $\mathbb{E}\{ f(x) \epsilon^*\} =0$ yields the right-hand side of \eqref{eq:Bussgang-expression}.
\end{IEEEproof}

Using Lemma~\ref{lemma:bussgang}, it follows that (cf.~\cite[Appendix~A]{Jacobsson2017a})
\begin{equation} \label{eq:Czu_expression}
\vect{C}_{zu} =\EcondH\{ \vect{z} \vect{u}^{\Htran} \}= \vect{D} \vect{C}_{uu}
\end{equation}
where $\vect{D} = \diag(d_1,\ldots,d_M)$ and $d_m = \frac{\EcondH\{ g_m(u_m) u_m^\star \}}{  \EcondH \{ | u_m |^2 \}}$. 
Inserting \eqref{eq:Czu_expression} into  \eqref{eq:z-distortion2} yields
\begin{equation} \label{eq:z-distortion3}
\vect{z} = \vect{D}\vect{u} + \boldsymbol{\eta}
\end{equation}
with $\boldsymbol{\eta} = \boldsymbol{g}(\vect{u}) - \vect{D}\vect{u}$. The following observation 
is thus made.

\begin{observation}
When a Gaussian signal $\vect{u}$ is affected by non-ideal BS receiver hardware, the output is an element-wise scaled version of $\vect{u}$ plus a distortion term $\boldsymbol{\eta}$ that is uncorrelated with $\vect{u}$.
\end{observation}

If the functions $g_m(\cdot)$ are all equal and $\vect{C}_{uu}$ has identical diagonal elements, $\vect{D}$ is a scaled identity matrix and the common scaling factor represents the power-loss incurred by hardware impairments. In this case, $ \vect{D}\vect{u}$ has the same correlation characteristics as $\vect{u}$. In other cases, one can calibrate the hardware to make $ \vect{D}$ a scaled identity matrix.

The distortion term has the (conditional) correlation matrix
\begin{equation} \label{eq:Cetaeta}
\vect{C}_{\eta \eta} = \EcondH \{\boldsymbol{\eta}\boldsymbol{\eta}^{\Htran}\} = \vect{C}_{zz} - \vect{C}_{zu} \vect{C}_{uu}^{\dagger} \vect{C}_{zu}^{\Htran} =  \vect{C}_{zz} - \vect{D} \vect{C}_{uu} \vect{D}^{\Htran}
\end{equation}
where $ \vect{C}_{zz} = \EcondH \{ \vect{z} \vect{z}^{\Htran} \}$. In the special case when $\vect{C}_{uu}$ is diagonal and $\EcondH\{ g_m(u_m) \} = 0$ for $m=1,\ldots,M$, $\vect{C}_{\eta \eta}$ is also diagonal. Hence, the distortion term has uncorrelated elements. This cannot happen unless $K \geq M$.

\begin{observation}
The elements of the BS distortion term $\boldsymbol{\eta}$ are generally correlated.
\end{observation}

This confirms previous downlink \cite{Moghadam2012a,Mollen2018b,Larsson2018a} and uplink \cite{Mollen2018a} results, derived with different system models. The question is now if the simplifying assumption of uncorrelated distortion (e.g., made in \cite{Bai2013a,Orhan2015a,Bjornson2014a}) has significant impact on the SE. 

\subsection{Spectral Efficiency with BS Hardware Impairments}

Using the signal and distortion characteristics derived above, we can determine the communication performance. The signal detection is based on the received signal $\vect{y}  \in \mathbb{C}^{M}$ that is available in the digital baseband. It is assumed to be
\begin{align} 
\vect{y} = \vect{z} + \vect{n} =  \vect{D}\vect{u} + \boldsymbol{\eta} + \vect{n}  =  \sum_{k=1}^{K}\vect{D}\vect{h}_k s_k + \boldsymbol{\eta} + \vect{n}   \label{eq:distorted-received-signal}
\end{align}
where $\vect{n} \sim \CN(\vect{0},\sigma^2 \vect{I}_M)$ accounts for thermal noise that is (conditionally) uncorrelated\footnote{In practice, the initially independent noise at the BS is also distorted by the BS hardware, which results in uncorrelated noise (conditioned on a channel realization $\vect{H}$) by following the same procedure as above.} with $\vect{u}$ and $\boldsymbol{\eta}$. The combining vector $\vect{v}_k$ is used to detect the signal of UE $k$ as
\begin{align}
\vect{v}_k^{\Htran}\vect{y} = \vect{v}_k^{\Htran} \vect{D}\vect{h}_k s_k +   \sum_{i=1, i\ne k}^{K} \vect{v}_k^{\Htran} \vect{D}\vect{h}_i s_i + \vect{v}_k^{\Htran}\boldsymbol{\eta} + \vect{v}_k^{\Htran}\vect{n}   \label{eq:distorted-received-signal2}.
\end{align}
In the given coherence block, $\boldsymbol{\eta}$ is uncorrelated with $\vect{u}$, thus the distortion is also uncorrelated with the information-bearing signals $s_1,\ldots,s_K$ (which are mutually independent by assumption).
Hence, we can use the \emph{worst-case uncorrelated additive noise theorem} \cite{Hassibi2003a} to lower bound the mutual information between the input $s_k$ and output $\vect{v}_k^{\Htran}\vect{y}$ in \eqref{eq:distorted-received-signal2} as \vspace{-1mm}
\begin{align}  
\mathcal{I}(s_k; \vect{v}_k^{\Htran}\vect{y}) \geq \log_2 \left( 1 + \gamma_k \right) \label{eq:first-SE-bound}
\end{align}
for the given deterministic channel realization $\vect{H}$, where $\gamma_k $ represents the instantaneous signal-to-interference-and-noise ratio (SINR) and is given by
\begin{align}\label{eq:SINR}
\gamma_k = \frac{ p \vect{v}_k^{\Htran} \vect{D}\vect{h}_k \vect{h}_k^{\Htran} \vect{D}^{\Htran}  \vect{v}_k}{  \vect{v}_k^{\Htran} \big(\sum\limits_{i \neq k} p \vect{D}\vect{h}_i \vect{h}_i^{\Htran} \vect{D}^{\Htran} + \vect{C}_{\eta \eta}  +  \sigma^2 \vect{I}_M \big)  \vect{v}_k}.
\end{align}
Since $\gamma_k$ is a generalized Rayleigh quotient with respect to $\vect{v}_k$, it is maximized by \cite[Lemma B.10]{massivemimobook}
\begin{equation} \label{eq:combining-vector}
\vect{v}_k = p \bigg( \sum_{i=1,i \neq k}^K p \vect{D}\vect{h}_i \vect{h}_i^{\Htran} \vect{D}^{\Htran} \!+\! \vect{C}_{\eta \eta} \! +\!  \sigma^2 \vect{I}_M \bigg)^{\!-1}  \vect{D} \vect{h}_k.
\end{equation}
We call this the \emph{distortion-aware minimum-mean squared error} (DA-MMSE) receiver as it takes into account not only inter-user interference and noise, but also the distortion correlation.

\begin{observation}
The  BS distortion correlation affects the SINR and can be utilized in the receive combining. The SE-maximizing combining vector is changed by the correlation.
\end{observation}

Substituting \eqref{eq:combining-vector} into \eqref{eq:SINR} yields
\begin{align} 
\!\!\!\!\gamma_k= p \vect{h}_k^{\Htran} \vect{D}^{\Htran}  \bigg( \!\sum_{i=1,i \neq k}^K \!\!\!p \vect{D}\vect{h}_i \vect{h}_i^{\Htran} \vect{D}^{\Htran} \!+\! \vect{C}_{\eta \eta} \! +\!  \sigma^2 \vect{I}_M \bigg)^{\!-1}  \!\!\!\! \vect{D} \vect{h}_k\!\!
\end{align}
so that the ergodic SE $\mathcal{I}(s_k; \vect{v}_k^{\Htran}\vect{y}, \vect{H}) = \mathbb{E}_{\vect{H}}\{\mathcal{I}(s_k; \vect{v}_k^{\Htran}\vect{y}) \} $ over the fading channel in \eqref{eq:distorted-received-signal}, where $\mathbb{E}_{\vect{H}}$ denotes expectation w.r.t.~$\vect{H}$, is lower bounded by \vspace{-1mm}
\begin{equation} 
\mathbb{E}_{\vect{H}}\{\mathcal{I}(s_k; \vect{v}_k^{\Htran}\vect{y}) \} \ge \mathbb{E}_{\vect{H}} \{ \log_2 ( 1+ \gamma_k) \}. \label{SE-ideal-tx}
\end{equation}

\subsection{Spectral Efficiency with UE Hardware Impairments}
\label{subsec:SE-non-ideal-tx}

In practice, there are hardware impairments at both the BS and UEs. To quantify the relative impact of both impairments, we next assume that $s_k = \varsigma_k + \omega_k$ for $k=1,\ldots,K$, where $\varsigma_k \sim \CN(0, \kappa p)$ is the actual desired signal from UE $k$ and $\omega_k \sim \CN(0, (1-\kappa) p)$ is a distortion term. The parameter $\kappa \in [0,1]$ determines the level of hardware impairments at the UE side, potentially after predistortion. For analytical tractability, we assume that $\varsigma_k$ and $\omega_k$ are independent, thus the transmit power is $\mathbb{E}\{ |s_k|^2\} = \kappa p + (1-\kappa) p = p$ irrespective of $\kappa$. The independence is a worst-case assumption \cite{Hassibi2003a,massivemimobook}, but is mainly made to obtain the achievable SE $\mathbb{E}_{\vect{H}}\{\mathcal{I}(\varsigma_k; \vect{v}_k^{\Htran}\vect{y}) \} \geq \mathbb{E}_{\vect{H}}\{ \log_2(1+\gamma_k^\prime)\}$ with (using the same methodology as above) \vspace{-6mm}
\begin{align} \notag 
&\gamma_k^\prime= \\ 
&\frac{ \kappa p \vect{v}_k^{\Htran} \vect{D}\vect{h}_k \vect{h}_k^{\Htran} \vect{D}^{\Htran}  \vect{v}_k}{  \vect{v}_k^{\Htran} \big( \! \sum\limits_{i \neq k} p \vect{D}\vect{h}_i \vect{h}_i^{\Htran} \vect{D}^{\Htran} \!+ p (1\!-\!\kappa)\vect{D}\vect{h}_k \vect{h}_k^{\Htran} \vect{D}^{\Htran}  \! + \vect{C}_{\eta \eta}  +  \sigma^2 \vect{I}_M \big)  \vect{v}_k}. \label{SE-non-ideal-tx}
\end{align}
This SINR is also maximized by DA-MMSE in \eqref{eq:combining-vector}, as it can be proved using \cite[Lemma B.4]{massivemimobook}. The reason is that the desired signal and UE distortion are received over the same channel $\vect{D}\vect{h}_k$, thus such distortion cannot be canceled by receive combining without canceling the desired signal. 

\begin{observation}
The UE distortion does not change the SE-maximizing receive combining vector at the BS.
\end{observation}

\section{Quantifying the Impact of Non-linearities}
\label{sec:non-linearities}

\begin{figure}
  \centering  \vspace{-2mm}
    \includegraphics[width=0.5\textwidth]{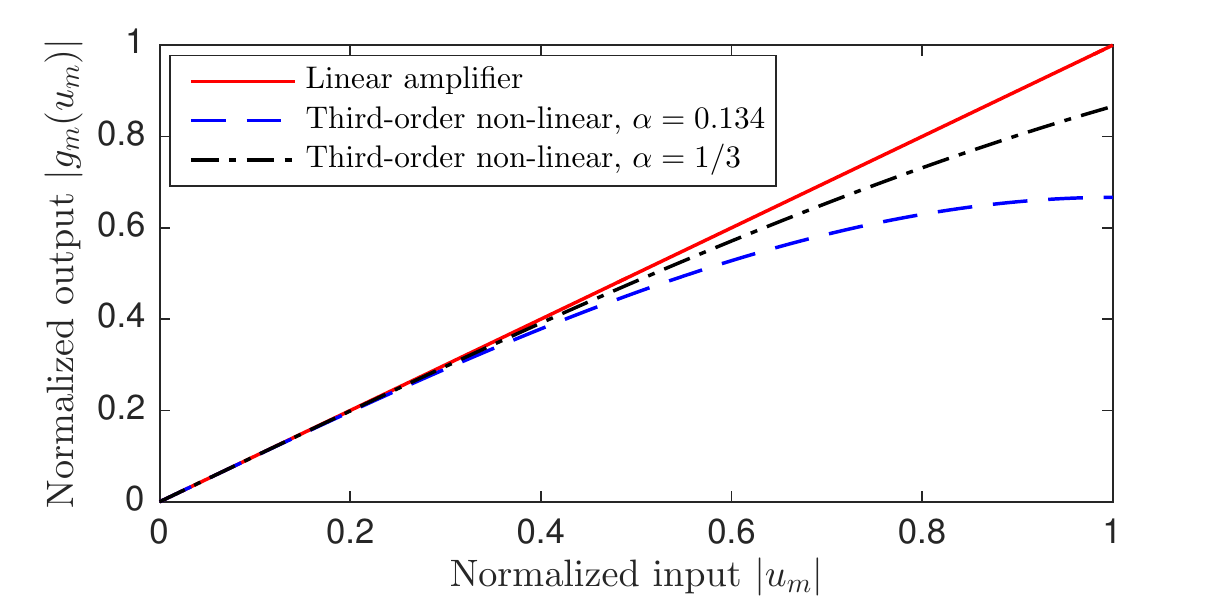} \vspace{-6mm}
      \caption{Comparison of linear and non-linear amplifiers in \eqref{eq:third-order-nonlinear} with $\boff=1$.} \label{figure_amplifier}  \vspace{-3mm}
\end{figure}

The distortion term $\vect{v}_k^{\Htran} \vect{C}_{\eta \eta}   \vect{v}_k$ appears in \eqref{eq:SINR} and \eqref{SE-non-ideal-tx}.
To analyze the characteristics of this term and, particularly, the impact of the distortion correlation (i.e., the off-diagonal elements in $\vect{C}_{\eta \eta}$), we consider the AM-AM distortion caused by the LNA. This can be modeled, in the complex baseband, by a third-order strictly memoryless non-linear function \cite{Schenk2008a}
\begin{equation} \label{eq:third-order-nonlinear}
g_m(u_m) = u_m - a_m |u_m|^2 u_m, \quad  m=1,\ldots,M.
\end{equation} 
This is a valid model of amplifier saturation when $a_m \geq 0$ and for such input amplitudes $|u_m|$ that $|g_m(u_m)|$ is an increasing function.
This occurs for $|u_m| \leq \frac{1}{\sqrt{3a_m}}$, while clipping occurs for input signals with larger amplitude. 
The value of $a_m$ depends on the circuit technology and how the output power of the LNA is normalized. One can model it as
\begin{equation} \label{eq:a_m_model}
a_m = \frac{\alpha}{\boff\mathbb{E}\{ |u_m|^2 \} }
\end{equation}
where $\mathbb{E}\{ |u_m|^2 \} $ is the average signal power and $\boff \geq 1$ is the back-off parameter selected based on the peak-to-average-power ratio (PAPR) of the input signal to limit the risk for clipping. The parameter $\alpha>0$ determines the non-linearities for normalized input signals with amplitudes in $[0,1]$. The worst case is given by $\alpha= 1/3$, for which the LNA saturates at unit input amplitude. A smaller value of $\alpha = 0.1340$ was reported in \cite{3GPP_PA_models} for a GaN amplifier operating at 2.1\,GHz. These amplifiers are illustrated in Fig.~\ref{figure_amplifier} for $\boff=1$. 

We can use this model to compute $\vect{C}_{\eta \eta}$. We let $\rho_{ij} = \EcondH\{ u_i u_j^*\} = [\vect{C}_{uu}]_{ij}$ denote the $ij$th element of $\vect{C}_{uu}$.
With this notation, $d_m = \frac{\rho_{mm} - 2 a_m \rho_{mm}^2}{\rho_{mm}}= 1 - 2 a_m \rho_{mm}$ and thus
\begin{equation} \label{eq:DCssD}
[\vect{D} \vect{C}_{uu} \vect{D}^{\Htran}]_{ij} = d_i \rho_{ij} d_j^* = (1 - 2 a_i \rho_{ii}) \rho_{ij} (1 - 2 a_j \rho_{jj}).
\end{equation}
The (conditional) correlation matrix of $\vect{z}$ has elements
\begin{align} \notag
& [\vect{C}_{zz} ]_{ij} = \EcondH\{ g_i(u_i) \left( g_j(u_j) \right)^*\} \\ \notag & =  \EcondH\{ u_i u_j^*\} - a_i \EcondH\{  |u_i|^2 u_i  u_j^* \}- a_j \EcondH\{ u_i |u_j|^2 u_j^* \}  
\\ \label{eq:derivationCzz1} & \quad + a_i a_j \EcondH\{ |u_i|^2 |u_j|^2 u_i u_j^* \} \\ \notag
&=  \rho_{ij} \!-\! 2  a_i  \rho_{ii} \rho_{ij} \!-\! 2  a_j \rho_{jj} \rho_{ij}  \!+\! a_i a_j (2 | \rho_{ij}|^2 \rho_{ij} \!+\! 4 \rho_{ij} \rho_{ii} \rho_{jj} ) \notag \\
& = (1 - 2 a_i \rho_{ii}) \rho_{ij} (1 - 2 a_j \rho_{jj}) + 2 a_i a_j  | \rho_{ij}|^2  \rho_{ij} \label{eq:Czz}
 \end{align}
where the first expectation in \eqref{eq:derivationCzz1} is $\rho_{ij}$, as defined above, while the second and third expectations can be computed using Lemma~\ref{lemma:bussgang}. To compute the last expectation, we follow the procedure in the proof of Lemma~\ref{lemma:bussgang} to show that
$u_i = \frac{\rho_{ij}}{\rho_{jj}} u_j + \epsilon$, where $\epsilon \sim \CN (0,\rho_{ii}-{|\rho_{ij}|^2}/{\rho_{jj}} )$ is independent of $u_j$. Substituting this into the last expectation, it follows that
\begin{align} \notag
&\EcondH\{ |u_i|^2 |u_j|^2 u_i u_j^* \} \\ \notag &= \EcondH \left\{ \left|\frac{\rho_{ij}}{\rho_{jj}} u_j + \epsilon \right|^2 \! |u_j|^2 \left(\frac{\rho_{ij}}{\rho_{jj}} u_j + \epsilon\right) u_j^* \right\} \\ \notag
& = \left|\frac{\rho_{ij}}{\rho_{jj}}  \right|^2 \frac{\rho_{ij}}{\rho_{jj}} \EcondH \{ |u_j|^6  \} + 2 \frac{\rho_{ij}}{\rho_{jj}}  \EcondH \{|u_j|^4 \} \EcondH\{ |\epsilon |^2\} \\ 
& = 2 | \rho_{ij}|^2 \rho_{ij} + 4 \rho_{ij} \rho_{ii} \rho_{jj}.
\end{align}
By using \eqref{eq:DCssD} and \eqref{eq:Czz}, we can now obtain the elements of the distortion term's correlation matrix in \eqref{eq:Cetaeta} as
\begin{align} \label{eq:Cetaeta_expression1}
 [\vect{C}_{\eta \eta} ]_{ij} =  [\vect{C}_{zz} ]_{ij} - [\vect{D} \vect{C}_{uu} \vect{D}^{\Htran}]_{ij} = 2 a_i a_j  | \rho_{ij}|^2  \rho_{ij}
\end{align}
which can be expressed in matrix form as 
\begin{align} \label{eq:Cetaeta_expression}
\vect{C}_{\eta \eta} = 2 \vect{A} \left( \vect{C}_{uu} \odot \vect{C}_{uu}^* \odot \vect{C}_{uu}  \right) \vect{A}
\end{align}
where $\odot$ denotes the Hadamard (element-wise) product and $\vect{A} = \diag(a_1,\ldots,a_M)$.
If the information signal $\vect{u}$ has correlated elements (i.e., $ \vect{C}_{uu}$ has non-zero off-diagonal elements), it follows from \eqref{eq:Cetaeta_expression} that also the distortion has correlated elements. The correlation coefficient between $u_i$ and $u_j$ is
\begin{equation}
\xi_{u_i u_j} = \frac{\EcondH\{ u_i u_j^*\} }{\sqrt{\EcondH\{ |u_i|^2\}  \EcondH\{  |u_j|^2\} }} = \frac{\rho_{ij}}{\sqrt{\rho_{ii} \rho_{jj}}}
\end{equation}
while the correlation coefficient between $\eta_i$ and $\eta_j$ is
\begin{equation}
\xi_{\eta_i \eta_j} =\! \frac{\EcondH\{ \eta_i \eta_j^*\} }{\sqrt{\EcondH\{ |\eta_i|^2\}  \EcondH\{  |\eta_j|^2\} }} = \! \frac{|\rho_{ij}|^2 \rho_{ij}}{\sqrt{\rho_{ii}^3 \rho_{jj}^3}} = |\xi_{u_i u_j} |^2 \xi_{u_i u_j}.
\end{equation}
Clearly, $|\xi_{\eta_i \eta_j} | = |\xi_{u_i u_j} |^3 \leq |\xi_{u_i u_j} |$ since $ |\xi_{u_i u_j} | \in [0,1]$.

\begin{observation}
The distortion terms are less correlated than the corresponding signal terms.
\end{observation}

This is in line with observations made in~\cite{Moghadam2012a,Mollen2018b,Mollen2018a}.

\subsection{What Happens if the Distortion Correlation is Neglected?}

If the distortion terms are only weakly correlated, it would be analytically tractable to neglect the correlation. This effectively means using the diagonal correlation matrix
\begin{equation} \label{Cee_diag}
\vect{C}_{\eta \eta}^{\diag} = \vect{C}_{\eta \eta}  \odot \vect{I}_M
\end{equation}
which has the same diagonal elements as $ \vect{C}_{\eta \eta}$. This simplification is made in numerous papers that analyze SE \cite{Bai2013a,Orhan2015a,Bjornson2014a,massivemimobook}. We will now quantify the impact that such a simplification has when the BS distortion is caused by the third-order non-linearity in \eqref{eq:third-order-nonlinear}. 
For this purpose, we consider i.i.d.~Rayleigh fading channels $\vect{h}_k \sim \CN (\vect{0},\vect{I}_M)$ for $k=1,\ldots,K$. The average power received at BS antenna $m$ in \eqref{eq:a_m_model} is
\begin{equation}
\mathbb{E}\{ |u_m|^2 \} = \mathbb{E} \left\{ p \sum_{k=1}^{K} | h_{km} |^2 \right\} = p K.
\end{equation}
The impact of distortion correlation can be quantified by considering the distortion term $\vect{v}_k^{\Htran}\vect{C}_{\eta \eta} \vect{v}_k$   in \eqref{eq:SINR} and \eqref{SE-non-ideal-tx} and comparing it with $\vect{v}_k^{\Htran}\vect{C}_{\eta \eta}^{\diag} \vect{v}_k$ where correlation is neglected. To make a fair comparison, we consider MR combining with $\vect{v}_k = \vect{h}_k / \sqrt{\mathbb{E}\{ \| \vect{h}_k  \|^2\}}$, which does not  suppress distortion.

\begin{lemma} \label{lemma:ratios-of-distortion}
Consider i.i.d.~Rayleigh fading channels and $a_m$ given by  \eqref{eq:a_m_model},
then
\begin{align} \label{eq:ratio-of-distortion}
&\frac{\mathbb{E}\{ \vect{h}_k^{\Htran} \vect{C}_{\eta \eta} \vect{h}_k\}}{\mathbb{E}\{ \| \vect{h}_k \|^2 \} } = \frac{2 \alpha^2 p}{\boff
^2} \! \left( K+6+\frac{9}{K}+\frac{4}{K^2}+\frac{2M(K+1)}{K^2} \right)\! \\
\label{eq:ratio-of-distortion-uncorr}
& \approx \frac{\mathbb{E}\{ \vect{h}_k^{\Htran} \vect{C}_{\eta \eta}^{\diag} \vect{h}_k\}}{\mathbb{E}\{ \| \vect{h}_k \|^2 \} } = \frac{2 \alpha^2 p}{\boff^2} \! \left( K+6+\frac{11}{K}+\frac{6}{K^2} \right)
\end{align}
where the approximation neglects the distortion correlation.
\end{lemma}\vspace{-2mm}
\begin{IEEEproof}
Follows from direct computation of moments of complex Gaussian random variables.
\end{IEEEproof}

The average distortion power with MR combining is larger when the distortion terms are correlated, since the fraction
\begin{equation} \label{eq:distortion-fraction}
\!\!\!\frac{ \frac{\mathbb{E}\{ \vect{h}_k^{\Htran} \vect{C}_{\eta \eta} \vect{h}_k\}}{\mathbb{E}\{ \| \vect{h}_k \|^2 \} } }{ \frac{\mathbb{E}\{ \vect{h}_k^{\Htran} \vect{C}_{\eta \eta}^{\diag} \vect{h}_k\}}{\mathbb{E}\{ \| \vect{h}_k \|^2 \} } } = \frac{ \mathbb{E}\{ \vect{h}_k^{\Htran} \vect{C}_{\eta \eta} \vect{h}_k\} }{ \mathbb{E}\{ \vect{h}_k^{\Htran} \vect{C}_{\eta \eta}^{\diag} \vect{h}_k\} } 
= 1 + \frac{ 2(M-1) }{(K+2)(K+3)}\!\!
\end{equation}
is larger than one and independent of $\alpha$ and $\boff$. The size of the second term depends on the relation between $M$ and $K$.

\begin{figure}
  \centering \vspace{-2mm}
    \includegraphics[width=0.5\textwidth]{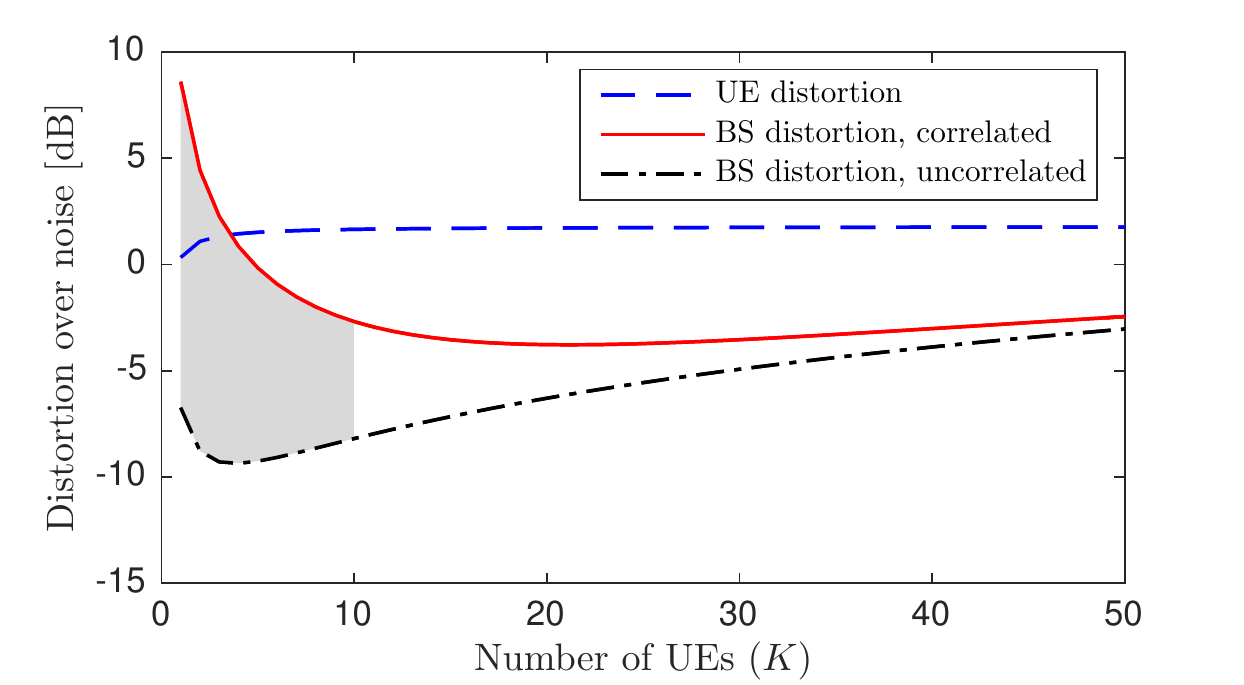} \vspace{-8mm}
      \caption{The BS and UE distortion power with correlation and when correlation is neglected. The approximation drops significantly in the shaded interval. The UE distortion dominates for $K\geq 5$.} \label{figure_distortioncomparison}  \vspace{-4mm}
\end{figure}

The solid and dash-dotted curves in Fig.~\ref{figure_distortioncomparison} show \eqref{eq:ratio-of-distortion} and \eqref{eq:ratio-of-distortion-uncorr}, normalized by the noise, as a function of $K$. We consider a Massive MIMO setup with $M=200$, a worst-case amplifier with $\alpha = 1/3$, $\boff = 7$\,dB, and a signal-to-noise ratio (SNR) of $p/\sigma^2=0$\,dB.
The correlation has a huge impact on the BS distortion term when there are few UEs. Quantitively speaking, it increases by more than 10\,dB. The reason is that the correlation gives the distortion vector $\boldsymbol{\eta}$ a similar direction as $\vect{h}_k$, for $k=1,\ldots,K$, when $K$ is small. Hence, the distortion effect is amplified by MR combining. The gap to the curve with uncorrelated distortion reduces with $K$. In the shaded part, the gap reduces from $15.3$ to $5.5$\,dB.
This is expected from \eqref{eq:distortion-fraction}, since $\boldsymbol{\eta}$ becomes less correlated with each UE signal that is added when $K$ grows. 

We now evaluate the impact of the distortion caused by non-ideal hardware at the UE. This results in $p (1-\kappa) | \vect{h}_k^{\Htran} \vect{D}^{\Htran} \vect{v}_k|^2$ in the denominator of the SINR in \eqref{SE-non-ideal-tx}. 
With i.i.d.~Rayleigh fading, MR combining, and  $a_m$ given by  \eqref{eq:a_m_model}, we have \vspace{-1mm}
\begin{align} \notag
&\frac{\mathbb{E}\{ | \vect{h}_k^{\Htran} \vect{D} \vect{h}_k |^2\}}{\mathbb{E}\{ \| \vect{h}_k \|^2 \} } = 
(M+1)-\frac{4\alpha (MK+K+M+3)}{\boff K} \\ &+\frac{4\alpha^2 (MK^2+8K+11+2MK+K^2+M)}{\boff^2 K^2}
\end{align}
which grows with $M$, similar to \eqref{eq:ratio-of-distortion}. The dashed curve in Fig.~\ref{figure_distortioncomparison} shows the UE distortion, under the same conditions as for the other curves. We consider high-quality transmitter hardware with $\kappa=0.99$ \cite[Sect.~6.1.2]{massivemimobook} and the signal-to-distortion power ratio $\kappa/(1-\kappa)=99$, which is higher than $[\vect{D} \vect{C}_{uu} \vect{D}^{\Htran}]_{ii} /  [\vect{C}_{\eta \eta} ]_{ii} \approx 85$  for the LNA. 
The correlated BS distortion is the dominant factor for $K \leq 3$.
But  for larger values of $K$ (as envisaged in Massive MIMO), the UE distortion becomes much higher (5\,dB in this example). The reason is that the largest BS distortion terms reduce with $K$.

\begin{observation}
The correlation of the BS distortion reduces with $K$. The BS distortion will eventually have a smaller impact than the UE distortion, which doesn't reduce with $K$.
\end{observation}

The first part of this observation is line with the downlink analysis in \cite{Moghadam2012a,Mollen2018b,Larsson2018a} and uplink analysis in \cite{Mollen2018a}. These papers did not quantify the impact of BS distortion on the SE.

\subsection{How Much is the SE Affected by Neglecting Distortion Correlation at the BS?}

The BS distortion is correlated among antennas, but (as shown in Fig.~\ref{figure_distortioncomparison}) this has a very limited impact on the total distortion in the SINR for $K \geq 5$. To further quantify the impact, we compute the SE expressions derived in Section~\ref{subsec:SE-non-ideal-tx} numerically for the same setup as considered in Fig.~\ref{figure_distortioncomparison}: $M=200$ antennas, varying number of UEs, i.i.d.~Rayleigh fading, $p/\sigma^2=0$\,dB SNR, and non-ideal hardware at the BS and UEs represented by $\alpha = 1/3$, $\kappa=0.99$, and $\boff=7$\,dB.

\begin{figure}
  \centering  \vspace{-2mm}
    \includegraphics[width=0.5\textwidth]{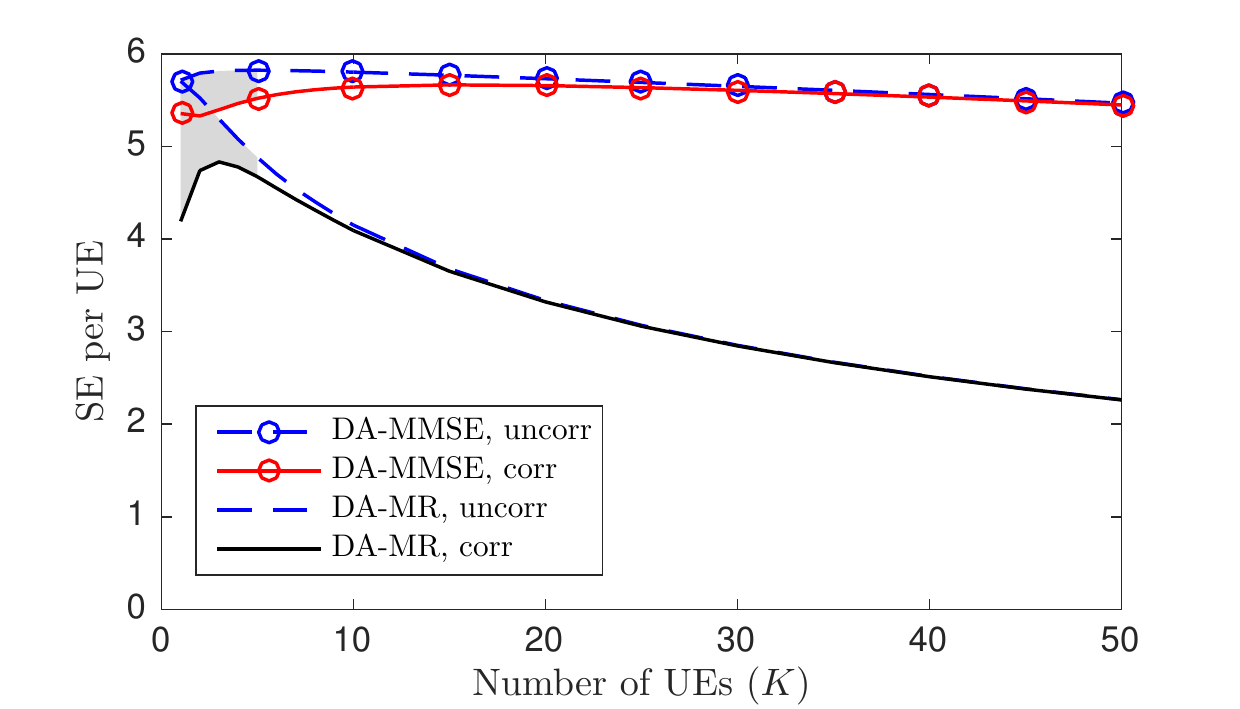} \vspace{-5mm}
      \caption{SE per UE when correlation of the BS distortion is neglected or accounted for with DA-MMSE and DA-MR.} \label{figure_SEdifference}  \vspace{-3mm}
\end{figure}

Fig.~\ref{figure_SEdifference} shows the SE per UE, as a function of the number of UEs. We consider either DA-MMSE combining in \eqref{eq:combining-vector} or distortion-aware MR (DA-MR) combining, defined as $\vect{v}_k = \vect{D} \vect{h}_k/\| \vect{D} \vect{h}_k\|$. The solid lines represent the exact SE, taking the correlated distortion into account, while the dashed lines represent approximate SEs achieved by neglecting distortion correlation; that is, using $\vect{C}_{\eta \eta}^{\diag} $ in
\eqref{Cee_diag} instead of $\vect{C}_{\eta \eta}$. Although the choice of the receive combining scheme has a large impact on the SE, the approximation error is negligible for $K \geq 5$ with both schemes. Even in the range $K< 5$, the shaded gap only ranges from 6.7\% to 5.5\% for DA-MMSE.

\begin{observation}
The distortion correlation has negligible impact on the uplink SE in the studied Massive MIMO scenario; that is, i.i.d.~Rayleigh fading and equal SNRs for all UEs.
\end{observation}

This is in line with the Massive MIMO papers \cite{Bjornson2014a,massivemimobook}, which make similar claims but without analytical or numerical evidence. Note that we reached this result by considering higher hardware quality at the UE than at the BS, in an effort to not underestimate the impact of BS distortion. In practice, the UEs could potentially have lower hardware quality than the BS, which would make the approximation error negligible for even smaller values of $K$. Furthermore, the use of a flat-fading channel, AM-AM distortion without phase-distortion, and no quantization errors are further conservative assumptions.

\section{Conclusions}

The hardware distortion in a multiple-antenna BS is generally correlated across antennas. 
This reduces the SINR, but its impact on the SE is marginal, particularly for DA-MMSE. Even in Massive MIMO with 200 antennas, approximating the distortion as uncorrelated leads to negligible errors when serving a small number of UEs. The reason is that the BS distortion correlation reduces with the number of UEs, making UE distortion the dominant factor. This was demonstrated by deriving SE expressions with arbitrary quasi-memoryless distortion functions and then quantifying the impact of third-order AM-AM non-linearities. Results were obtained analytically and numerically for i.i.d.~Rayleigh fading with equal SNRs. 

Different phenomena may arise when using other propagation and hardware models. Frequency-selective fading lead to reduced correlation \cite{Mollen2018b}. Distortion compensation algorithms achieve similar results. However, near-far effects might make the distortion vector similar to a cell-center UE's channel \cite{Larsson2018a}. 

In summary, we have demonstrated that the uncorrelated distortion model advocated in  \cite{Bjornson2014a,massivemimobook} for Massive MIMO with Rayleigh fading can give very accurate results when analyzing the SE in such systems. Although it is generally ``{\emph{physically inaccurate}}'' to neglect distortion correlation, we can do it when analyzing the SE. However, when using the model in other setups, one must always verify that the distortion correlation has negligible impact also in those setups.


\bibliographystyle{IEEEtran}
\bibliography{IEEEabrv,refs.bib}

\end{document}